\documentclass{article}

\usepackage{arxiv}

\usepackage[utf8]{inputenc} 
\usepackage[T1]{fontenc}    
\usepackage[hidelinks]{hyperref}       
\usepackage{url}            
\usepackage{booktabs}       
\usepackage{amsfonts}       
\usepackage{nicefrac}       
\usepackage{microtype}      
\usepackage{lipsum}		

\usepackage[super]{natbib}
\usepackage{graphicx}
\usepackage{gensymb}
\usepackage{siunitx}
\usepackage{caption}
\usepackage{notes2bib}
\usepackage{amsmath}
\usepackage{mciteplus}
\usepackage{listings}
\usepackage{bibunits}
\setcitestyle{citesep={,}}
\defaultbibliographystyle{unsrt}

\newcommand{\angstrom}{\textup{\AA}}

\title{Extending generalized Debye analysis to long timescale magnetic relaxation}

\author{
  Jeremy D. Hilgar \\
  Department of Chemistry and Biochemistry\\
  University of California, San Diego\\
  La Jolla, California \\
   \And
 Aaron K. Butts \\
  Department of Chemistry and Biochemistry\\
  University of California, San Diego\\
  La Jolla, California \\
  \And
 Jeffrey D. Rinehart* \\
  Department of Chemistry and Biochemistry\\
  University of California, San Diego\\
  La Jolla, California \\
  \texttt{jrinehart@ucsd.edu} \\
}

\begin{document}
\maketitle

\begin{abstract}
As the ability to generate magnetic anisotropy in molecular materials continues to hit new milestones, concerted effort has shifted towards understanding, and potentially controlling, the mechanisms of magnetic relaxation across a large time and temperature space. Slow magnetic relaxation in molecules is highly temperature-, field-, and environment-dependent with the relevant timescale easily traversing ten orders of magnitude for current single-molecule magnets (SMM). The prospect of synthetic control over the nature of (and transition probabilities between) magnetic states make unraveling the underlying mechanisms an important yet daunting challenge. Currently, instrumental considerations dictate that the characteristic relaxation time, $\tau$, is determined by separate methods depending on the timescale of interest. Static and dynamic probe fields are used for long and short timescales, respectively. Each method captures a distinct, non-overlapping time range, and experimental differences lead to the possibility of fundamentally different meanings for $\tau$ being plotted and fitted globally as a function of temperature. Herein, we present a method to generate long-timescale waveforms with standard vibrating sample magnetometry (VSM) instrumentation, allowing extension of alternating current (AC) magnetic impedance measurements to SMMs and other superparamagnets with arbitrarily long relaxation time.
\end{abstract}


\begin{bibunit}
\section*{Introduction}
Single-molecule magnets (SMMs) are a class of zero-dimensional  materials that utilize the spin-orbit interaction to generate an axially anisotropic angular momentum. Due to a preferred orientation along the anisotropy axis, the magnetization vector exhibits slow relaxation, as if it were a superparamagnet with a classic double well potential replaced by discrete quantum states.\cite{sessoliHighspinMoleculesMn12O121993, sessoliMagneticBistabilityMetalion1993} As a form of superparamagnetism, single-molecule magnetism cannot result in magnetic ordering. In recent years, however, the timescale of magnetic relaxation has extended drastically such that single magnetic ions are now shown to retain magnetization information for timescales exceeding 1 s in the liquid nitrogen temperature regime. At temperatures well below the superparamagnetic blocking temperature, \textit{T}\textsubscript{B}, these highly anisotropic systems can exhibit "frozen" dynamics due to their extremely long relaxation times, allowing the performance of typical permanent magnet characterization experiments, such as magnetization \textit{vs.} field measurements that display hysteretic behavior despite the lack of a permanent magnetic ground state. Magnetization vs. field scans contain information about the field dependence of quantum tunneling as well as the maximal magnetization and coercivity of the material. They do not, however, contain quantitative information about the number and time constants of the various relaxation processes contributing to the magnetic relaxation. This time-domain data is readily obtained as phase-dependent AC magnetic susceptibilities between approximately 0.001 and 1 s (AC relaxation), yet its collection becomes problematic at longer timescales. The standard method for determining long timescale relaxation time constants (DC relaxation) is to apply a large static magnetic field to magnetize a macroscopic amount of microcrystalline SMM material, remove the field as quickly as possible, and track the decay towards zero magnetization as a function of time. These data are a valuable source of information, yet have several drawbacks that can obscure the nature of the isolated molecular zero-field relaxation. First, there is a significant gap in measurement timescale between that captured by AC and DC relaxation (generally 1--10 min., depending on instrumentation capability). Second, commercial instrumentation is generally only capable of setting zero field within $\pm$ 30 Oe, leading to an inherent decay offset. In very sensitive systems, this small field offset can drastically alter the time constant or even completely change the dominant magnetic relaxation mechanism. Third, the scan from high field to zero field can instigate the transfer of energy between the (ostensibly isolated) molecules leading to propagating, multi-molecular relaxation processes. This last point especially contributes to multi- and stretched exponential dynamics that can obscure the molecular zero-field dynamics of interest for comparison to AC susceptibility relaxation data. Given the influx of new long-timescale SMMs and the need for analysis tools that leverage existing instrumentation, we sought to devise a method to extend the dynamic magnetization measurement methodology of AC susceptibility to longer timescales. Importantly, this design required compatibility with commonly-employed research magnetometers, similarity to the input signal characteristics of commonly used AC waveforms for probing magnetic relaxation, and extendability of the measurement to arbitrarily-long relaxation regimes. Herein we present a technique fitting these criteria whereby phase-dependent magnetic susceptibilities are extracted from standard vibrating sample magnetometer (VSM) measurements. To provide an initial demonstration of the method's efficacy, we examine the relaxation behavior of a variant of the SMM "erbocene" sandwich motif, [K(18-c-6)][Er(hdcCOT)\textsubscript{2}] (\textbf{1}, hdcCOT = hexahydrodicyclopentacyclooctatetraenide dianion, 18-c-6 = 18-crown-6) that has not been previously magnetically characterized.  

\section*{Results and Discussion}

\subsection*{Synthesis and solid-state structure}
\begin{figure}
\centering
\includegraphics[width=0.4\linewidth]{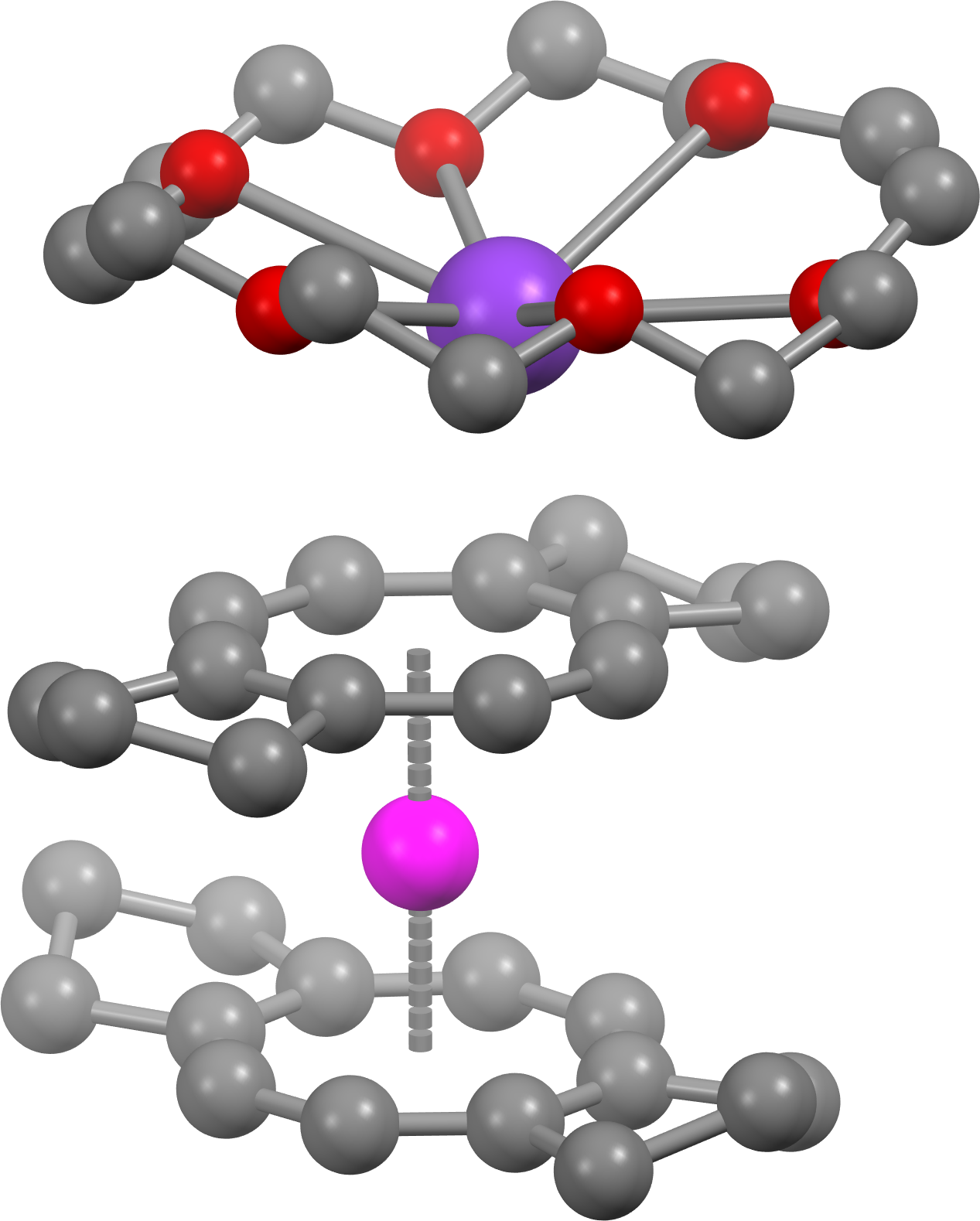}
\caption{Solid-state structure of \textbf{1} with spheres representing Er (pink), K (purple), O (red), and C (gray). Hydrogen atoms and lattice THF molecules have been omitted for clarity.}
\label{structure}
\end{figure}

Preparation of compound \textbf{1} was carried out under air-free conditions using a procedure similar to that employed for its previously reported lithio analog, [Li(THF)(DME)][Er(hdcCOT)\textsubscript{2}].\cite{hillerParamagneticNMRAnalysis2016a} Neutral hdcCOT was synthesized from 1,6-heptadiyne via a Ni\textsuperscript{0}-catalyzed dimerization reaction\cite{wenderNickelCatalyzedCycloadditions2007b} and was subsequently reduced using potassium graphite (KC\textsubscript{8}). The dipotassium salt of the alkyl-substituted COT dianion (K\textsubscript{2}hdcCOT) reacted readily with a suspension of erbium trichloride in THF to give K[Er(hdcCOT)\textsubscript{2}]; $\delta$H(500 MHz; THF-d\textsubscript{8}) --19.3 (br), --34.4 (br), --43.5 (br), --53.5 (br), --61.9 (br), --213.3 (br), --223.9 (br), --269.3 (br), and --343.3 (br) (Figures \ref{fig:nmr_full}--\ref{fig:nmr_up}). Removal of KCl and addition of 18-c-6 (to facilitate crystallization) yielded a solution which, when layered with pentane through vapor diffusion, deposited bright-yellow crystalline rods of \textbf{1}. X-ray analysis of these crystals showed that the solid-state structure of \textbf{1} includes an Er$^{3+}$ ion in a homoleptic coordination environment with two hdcCOT$^{2-}$ ligands coordinated in a 90\degree staggered conformation (Figure \ref{structure}). Steric bulk is presumably responsible for the observed conformation about the metal center, however this bulk imparts no appreciable elongation of the average Er-hdcCOT\textsubscript{centroid} distance relative to an unsubstituted analog (CCDC\cite{groom_cambridge_2016} Identifier YIWTUV\cite{meihausMagneticBlocking102013, ungurFinetuningLocalSymmetry2014}). The potassium ion in \textbf{1} supports a $\kappa{}^{6}$ interaction to 18-c-6 as well as an $\eta{}^{4}$ interaction to one face of a hdcCOT$^{2-}$ ring and the latter interaction lowers the hdcCOT--Er--hdcCOT sandwich angle from linearity to 178\degree.

\subsection*{Ab initio electronic structure}
Previous studies on [Er(COT)\textsubscript{2}]\textsuperscript{--} \cite{meihausMagneticBlocking102013, ungurFinetuningLocalSymmetry2014, leroySandwichComplexAxial2014, leroyCouplingStrategiesEnhance2014} and [Er(COT)]\textsuperscript{+}\cite{jiangOrganometallicSingleIonMagnet2011, mengBoratabenzeneCyclooctatetraenylLanthanide2016, hilgarFerromagneticCouplingChloridebridged2017, hilgarMetalLigandPair2018, chenSoftPhosphorusAtom2018, heEnantiopureBenzamidinateCyclooctatetraene2018, hilgarMillionfoldRelaxationTime2019} motifs have demonstrated that a remarkably consistent single-ion anisotropy is generated at Er\textsuperscript{3+} when suitably equatorial coordination environments are selected. The efficacy by which the planar COT dianion in particular can stabilize prolate, high-moment $m_{J}$ states on Er\textsuperscript{3+} has been well-rationalized by varying levels of theory and herein we modelled magnetic properties of \textbf{1} using complete active space, self-consistent field (CASSCF) methods (Figure \ref{fig:barrier}). Decomposition of the calculated ground state spin-orbit wavefunctions into an $m_{J}$ basis reveals that the lowered symmetry around the Er\textsuperscript{3+} center in \textbf{1} relative to unsubstituted [Er(COT)\textsubscript{2}]\textsuperscript{--} species plays a negligible role toward wavefunction mixing and consequently the ground Kramers doublet is highly axial with 97.7\% $m_{J} = \pm{}15/2$ character. The first two excited doublets (94.7\% $m_{J} =\pm{}13/2$ and 93.1\% $m_{J} = \pm{}1/2$) are predicted to lie 169.2 and 211.3 cm\textsuperscript{--1} above the ground doublet. Ground to first and second excited Kramers doublet splittings have previously been calculated for [K(18-c-6)][Er(COT)\textsubscript{2}]\cite{ungurFinetuningLocalSymmetry2014}; interestingly, we observe no appreciable difference in calculated splittings between \textbf{1} and its unsubstituted analog and we anticipate alkyl-substitutions on COT play a marginal role in affecting low-energy spin-orbit states on erbium. Transverse magnetic moment matrix elements connecting states within the $J = 15/2$ manifold have also been calculated. As discussed elsewhere,\cite{ungurInterplayStronglyAnisotropic2013} the square of these matrix elements are roughly proportional to transition rates between the states that they connect and the effective barrier of magnetic reversal ($U_{eff}$) in the high-temperature limit can be approximated as the shortest path between ground states with non-negligible matrix elements. Calculated matrix elements between the ground doublet and first excited doublet are small (3.3$\times$10\textsuperscript{--6} and 5.1$\times$10\textsuperscript{--4} $\mu_{B}$, respectively) which indicate that QTM between these states should be suppressed. Matrix elements connecting the second excited doublets ($m_{J} \sim \pm{}1/2$) and cross-terms between the first and second excited doublets (e.g. $m_{J} \sim +13/2$ and $m_{J} \sim -1/2$) are larger (3.2 and 4.1$\times$10\textsuperscript{--2} $\mu_{B}$, respectively) and likely relaxation pathways in the high temperature limit will involve thermally assisted quantum tunneling of the magnetization (QTM) or Orbach relaxation modes through the second excited Kramers doublet.

\subsection*{Magnetic characterization}
\subsubsection*{Static magnetism}
Zero-field cooled static magnetic susceptibilities for \textbf{1} were measured over 2--300 K with a 1000 Oe bias field and the data were plotted as the molar susceptibility times temperature product ($\chi{}T$) \textit{vs.} temperature (Figure \ref{fig:susceptibility}). At 300 K, $\chi{}T = 11.69$ emu mol$^{-1}$ K, which is close to the expected value for an Er\textsuperscript{3+} ensemble with equal populations across the $J = 15/2$ spin-orbit manifold ($\chi{}T = 11.49$ emu mol$^{-1}$ K). As $T$ is lowered from 300 K, $\chi{}T$ shows a small decline as this manifold becomes thermally depopulated. At 12 K, $\chi{}T = 11.24$ emu mol$^{-1}$ K; below this temperature the data display a precipitous drop to 1.60 emu mol$^{-1}$ K at 2 K. This drop is indicative of magnetic blocking on the timescale of the DC scan and thus isothermal magnetization measurements ($H_{max} = |7|$ T, $~10$ Oe sec\textsuperscript{--1} ramp rate) were collected to further probe for superparamagnetic behavior. At 2 K, the magnetization saturates at 4.98 $\mu_{B}$ mol\textsuperscript{--1} when the field is swept above 3.5 T. Consistent with the ZFC results, hysteretic behavior is observed when the field is swept back from 7 T. Near 0 T, the magnetization drops abruptly to 1.25 $\mu_{B}$ mol\textsuperscript{--1} and the application of negative fields yields a coercivity $H_{C}$ = 1.1 T on this timescale. Zero-field magnetization loss and waist-restricted hysteresis are commonly observed in SMMs when transverse crystal field (CF) components enable fast relaxation via QTM between ground states.\cite{liuSymmetryStrategiesHigh2018} Symmetry optimization strategies can be employed to mitigate this effect and in the case of [Er(COT)\textsubscript{2}]\textsuperscript{--} species the psuedo-C\textsubscript{$\infty$} axis minimizes transverse CF components to yield a ground doublet within which QTM is largely quenched. The striking zero-field magnetization loss in \textbf{1} mimics well the behavior observed in [K(18-c-6)][Er(COT)\textsubscript{2}] which, through a magnetic dilution study with yttrium, was shown to be facilitated by a bulk magnetic avalanche effect.\cite{meihausMagneticBlocking102013}
\subsubsection*{Dynamic magnetism}
\begin{figure}
\centering
\includegraphics[width=0.5\linewidth]{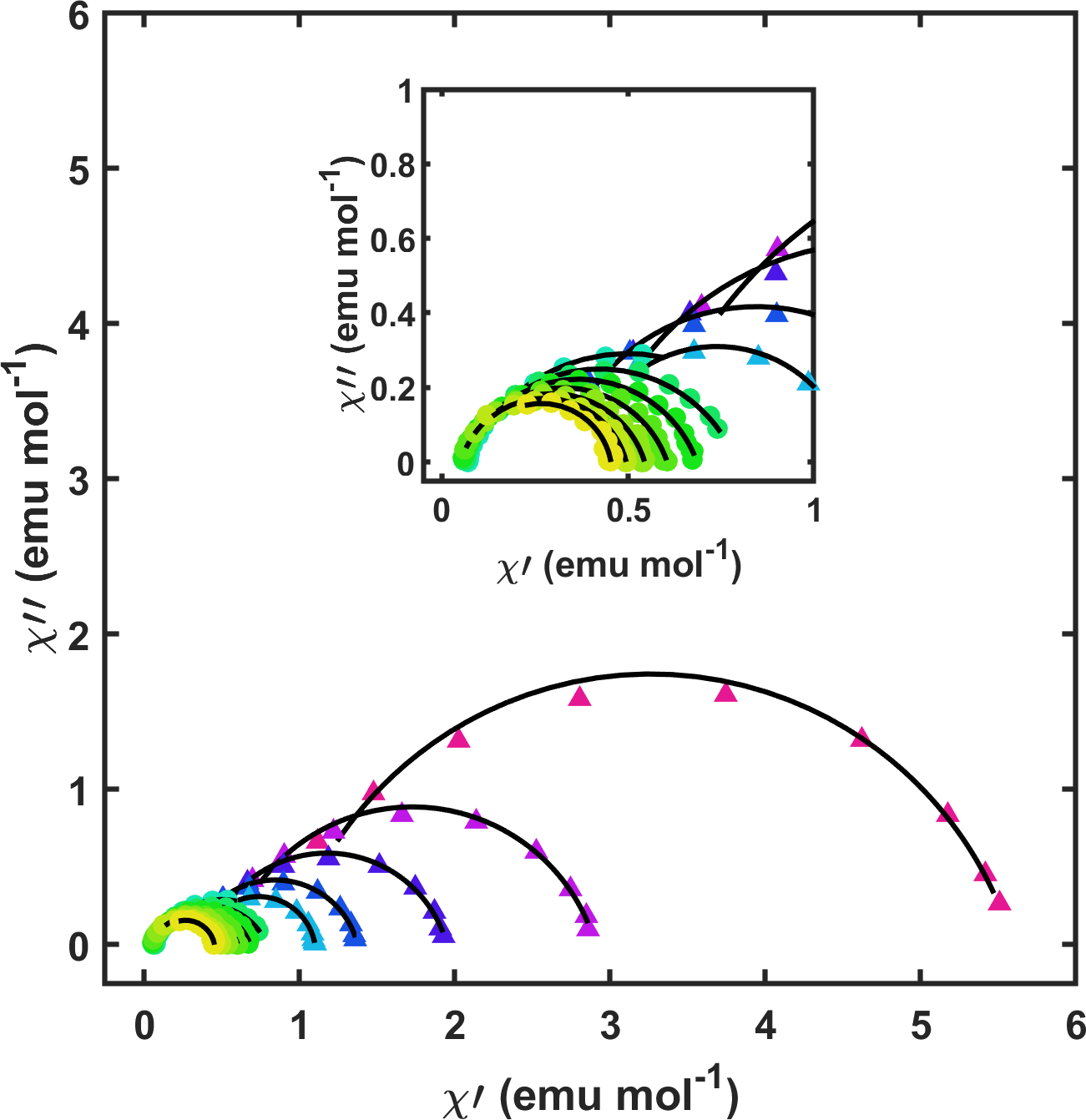}
\caption{Cole-Cole plot of \textbf{1} between 2--24 K. Colored points are susceptibilities measured via standard AC measurements (circles) and extracted from Fourier analysis of VSM data (triangles). Black lines represent fits a generalized Debye model (Equations 1 \& 2).}
\label{ac_combined}
\end{figure}

Consistent with SMM behavior, the AC susceptibility phase-shift of \textbf{1} shows a clear dependence on drive-field frequency below 24 K and short time-scale relaxation times were extracted from these data by simultaneously fitting the in-phase ($\chi{}\prime$) and out-of-phase ($\chi{}\prime{}\prime{}$) signals ($H_{DC}$ = 0 Oe, $f$ = 0.1--1000 Hz, $T$ = 12--24 K) to a generalized Debye relaxation model (Figure \ref{ac_combined}, Equations 1 and 2). \cite{gatteschiMolecularNanomagnets2006}
\begin{equation}
    \chi{}\prime{}(\omega{}) = \chi{}_{S} + (\chi{}_{T} - \chi{}_{S})\frac{1+(\omega{}\tau{})^{1-\alpha{}}sin(\pi{}\alpha{}/2)}{1+2(\omega{}\tau{})^{1-\alpha{}}sin(\pi{}\alpha{}/2)+(\omega{}\tau{})^{2-2\alpha{}}}
\end{equation}
\begin{equation}
    \chi{}\prime{}\prime{}(\omega{}) = (\chi{}_{T} - \chi{}_{S})\frac{(\omega{}\tau{})^{1-\alpha{}}cos(\pi{}\alpha{}/2)}{1+2(\omega{}\tau{})^{1-\alpha{}}sin(\pi{}\alpha{}/2)+(\omega{}\tau{})^{2-2\alpha{}}}
\end{equation}
\noindent{}Fitted eccentricities are low ($\alpha{}_{max} = 0.25(1)$) indicating a single relaxation time with a narrow distribution is associated with each temperature across the measured frequency range. A plot of $ln(\tau{})$ vs. $1/T$ (Figure \ref{arrhenius}) shows that an over-barrier Orbach relaxation mechanism is operant in this temperature range and least-squares fitting to an Arrhenius Law yielded an effective barrier height $U_{eff}$ = 147.7(7) cm\textsuperscript{--1} ($\tau{}_{0}$ = 3.1(1)$\times$10\textsuperscript{--8} s). A previous analysis on [K(18-c-6)][Er(COT)\textsubscript{2}] yielded an effective barrier $U_{eff}$ = 147 cm\textsuperscript{--1}; this striking similarity further corroborates our \textit{ab initio} results which suggested that the alkyl substitutions on the COT ring play a negligible role toward modifying the crystal field environment around Er\textsuperscript{3+}. 

\begin{figure}
\centering
\includegraphics[width=\linewidth]{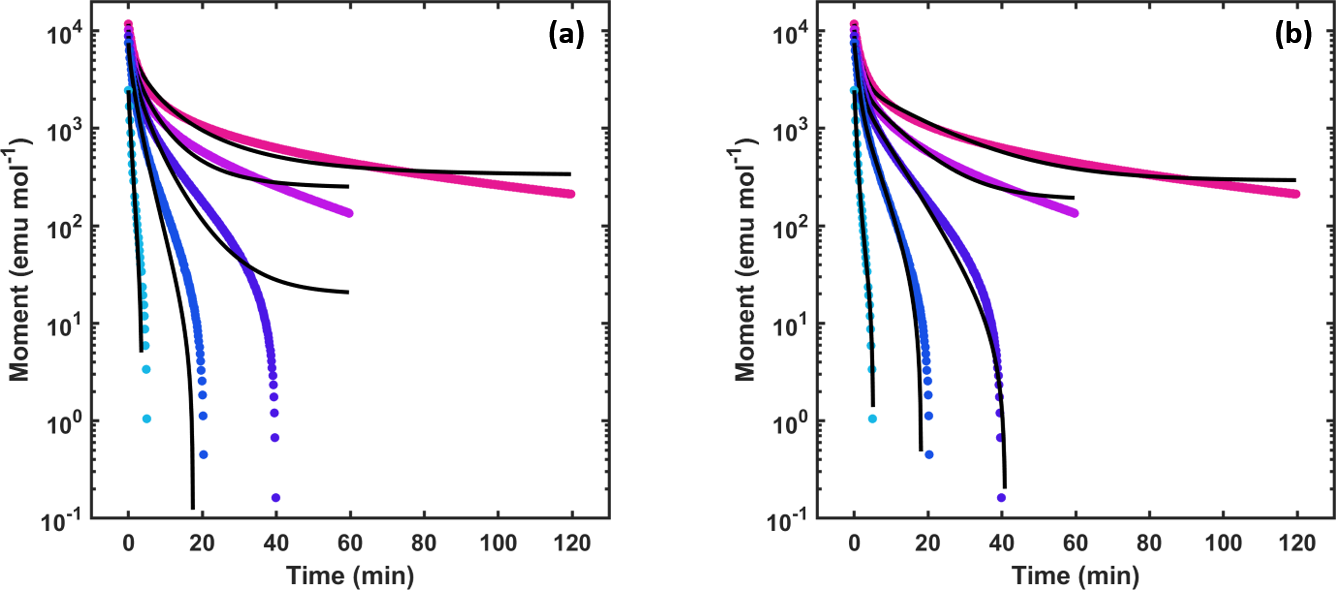}
\caption{DC relaxation data of \textbf{1} between 2--10 K. Colored points are moments measured from near-zero-field DC scans. Black lines are fits to (a) stretched and (b) bi-exponential models.}
\label{dc_combined}
\end{figure}

Preliminary studies on relaxation dynamics longer than the MPMS3 AC timescale ($< 0.1$ Hz) were conducted between 2--10 K by measuring the zero-field DC relaxation profile of a sample previously in equilibrium with a 7 T field. Consistent with the presence of a small remnant field in the MPMS3 superconducting magnet, final moments are negative and their magnitudes range from 0.37--1.8\% relative to the moment at $t$ = 0 s. This residual field is typically on the order of 20--30 Oe and two approaches used to minimize it include oscillating the field to zero and resetting the magnet (see e.g. "MPMS Application Note 1014-208," Quantum Design, 2001). We note that neither approach is feasible during the DC relaxation experiment. Given that multiple distinct processes or a distribution of processes are typically responsible for SMM relaxation at a given temperature, several forms of exponential decay models are routinely used to extract relaxation times in the long time-scale regime. Herein we fit DC relaxation data of \textbf{1} (Figure \ref{dc_combined}, Table 1) using a stretched exponential (Equation 3, 0 < $\beta{}$ < 1) and a bi-exponential model (Equation 4). 

\begin{equation}
M(t) = M_{f} + (M_{0} - M_{f})exp\left(\frac{-t}{\tau_{str}}\right)^{\beta}
\end{equation}
\begin{equation}
M(t) = M_{f} + (M_{0} - M_{f})\left[A_{1,2}exp\left(\frac{-t}{\tau_{1}}\right)+(1-A_{1,2})exp\left(\frac{-t}{\tau_{2}}\right)\right]
\end{equation}
Corroborating with the observation of multiple potential relaxation mechanisms in the magnetization data, we found that a simple exponential equation (Equation 3, $\beta{}$ = 1) does not adequately model the relaxation behavior of \textbf{1} below 10 K. Disagreement between data and fits become more prominent as temperature is lowered and it is evident that the extracted relaxation time is biased toward the short time-scale. Agreement between the data and fit can be improved with the addition of a stretching parameter, $\beta{}$ to the exponential. Similar to the $\alpha{}$ parameter of the generalized Debye equation, $\beta{}$ accounts for a distribution about an average relaxation process. Incorporation of $\beta{}$ results in the extraction of relaxation times with wide distributions at low T ($\beta{}$\textsubscript{2 K} = 0.505(9)). A similar quality fit was obtained when two exponential relaxation processes were used to model the data. Interestingly, this model reveals that relaxation of \textbf{1} can be rationalized by two processes operating on significantly different timescales (i.e. separated by an order of magnitude at 2 K) with the long relaxation time $\tau_{1}$ and short relaxation time $\tau_{2}$ bracketing the relaxation times obtained previously with a stretched exponential model. The presence of a small residual field (\textit{vide supra}) during the DC relaxation measurement provides a potential explanation for the observed long timescale process. This field can lift Kramers degeneracy between bistable orientations of the ground state and concomitantly hinder QTM. The addition of further constants and weighting factors would undoubtedly improve the fit, yet the ambiguity of the existing parameters and their importance to the intrinsic zero-field molecular relaxation make further parameterization of dubious value.   

\begin{table*}
\caption{Long time-scale relaxation times of \textbf{1} extracted from various models.}
\label{fit_values}
\centering
\begin{tabular}{@{\extracolsep{\fill}}
        *{8}{S[table-format=3.4]}}
\toprule
&\multicolumn{2}{c}{\textbf{Equations 1 \& 2}} & \multicolumn{2}{c}{\textbf{Equation 3}} & \multicolumn{3}{c}{\textbf{Equation 4}}\\
\cmidrule(lr){2-3} \cmidrule(lr){4-5} \cmidrule(lr){6-8}
{T (K)} & {$\tau_{VSM}$ (s)} & {$\alpha$} & {$\tau_{str}$ (s)} & {$\beta$} & {$\tau_{1}$ (s)} & {$\tau_{2}$ (s)} & {$A_{1,2}$}\\ 
\midrule
2 & 54(4) & 0.18(4) & 145(3) & 0.505(9) & 1093(29) & 87(1) & 0.222(4) \\ 
4 & 43(3) & 0.18(4) & 119(3) & 0.58(1) & 621(18) & 69(1) & 0.252(6) \\ 
6 & 37(3) & 0.17(4) & 106(2) & 0.62(1) & 442(9) & 61.8(8) & 0.271(5) \\ 
8 & 33(2) & 0.15(5) & 81(1) & 0.74(1) & 241(5) & 52.7(6) & 0.271(7) \\ 
10 & 13.4(8) & 0.10(3) & 36.3(4) & 0.88(1) & 122(8) & 30.6(4) & 0.13(1) \\ 
\bottomrule
\end{tabular} 
\\\footnotesize{Parentheses following fitted values represent 95\% confidence intervals.}
\end{table*}

\subsubsection*{VSM waveform analysis}
\begin{figure}
\centering
\includegraphics[width=\linewidth]{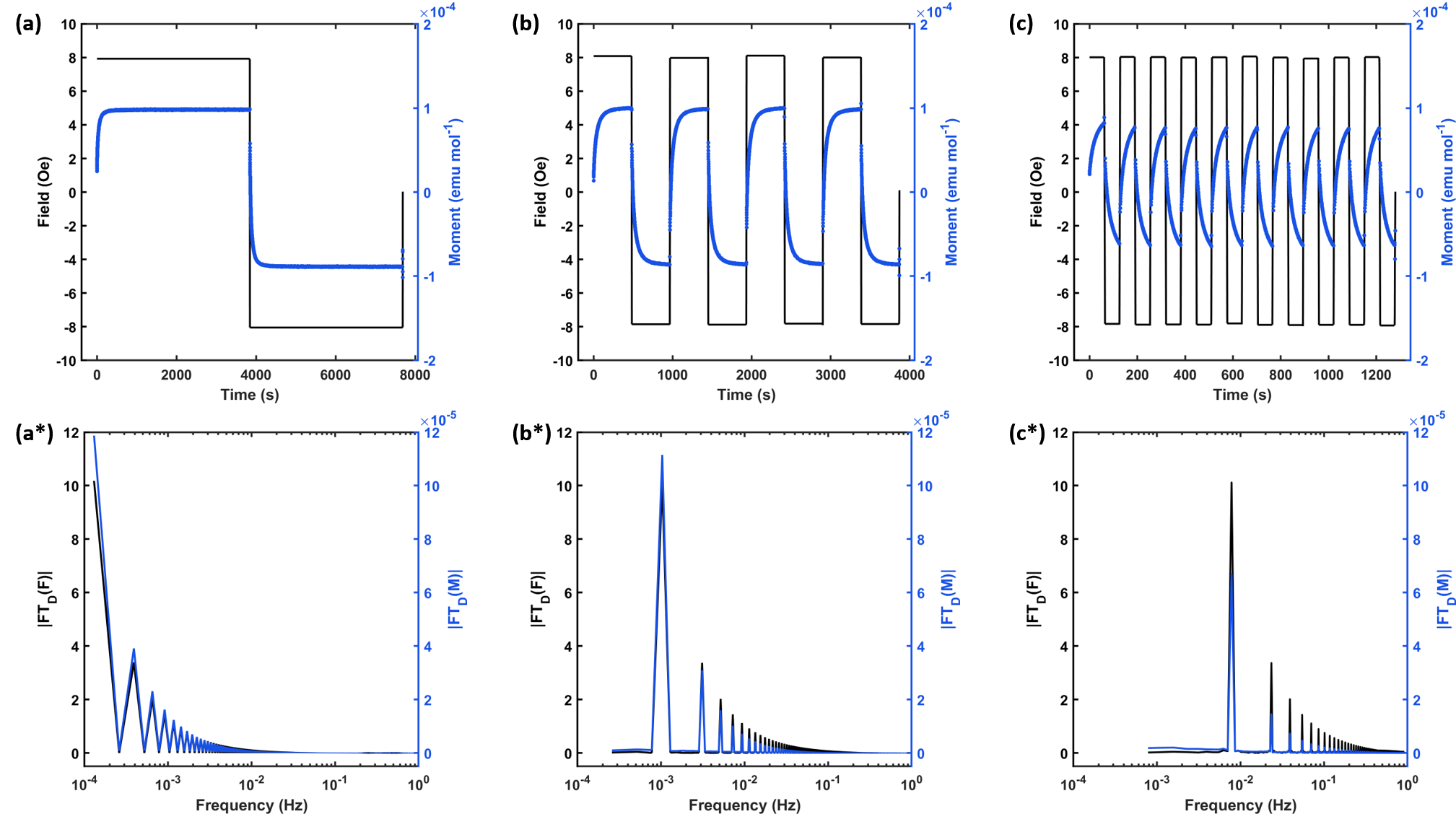}
\caption{Representative raw VSM waveform data (top row) and discrete Fourier transform of VSM waveform data (bottom row) with period lengths 7688 s (a and a*), 968 s (b and b*), and 128 s (c and c*). Black lines are field values and purple lines are measured moments.}
\label{newAC}
\end{figure}

Prompted by the desire for more quantitative information about long-timescale relaxation, we sought a method that could more faithfully model the zero-field relaxation dynamics of \textbf{1}. Ideally the method would be well-suited for extracting $\tau{}$ information from non-ensemble magnetic processes and would furthermore be capable of directly probing for the existence of multiple such relaxation processes. Naturally, Debye model fitting of complex AC magnetic susceptibilities fit these criteria well and is the standard when relaxation times fall within common magnet modulation coil limits ($\nu$ = 0.1--1500 Hz). An interesting question that follows from the preceding is whether complex susceptibilities can be reliably extracted from data collected from waveforms constructed in either DC or VSM scan modes. Although magnetic field controls have a limited linear sweep rate ($R_{H_{max}}$ = 1.6 Oe sec\textsuperscript{--1} for the Quantum Design MPMS3 magnetometer), drive fields of arbitrarily low frequency can be generated if an appropriate waveform is selected. To test whether phase information could be extracted from \textbf{1} outside the typical AC scan range, we developed a sequence which measures VSM moments generated by a square-wave magnetic drive field oscillating about 0 Oe ($H_{max}$ = 8 Oe, $R_{H}$ = 30 Oe sec\textsuperscript{--1}). Our initial studies on this approach yielded profiles of average moment significantly different from 0 emu mol\textsuperscript{--1}, indicating a persistent offset from zero bias field. A single magnet reset operation before data collection was sufficient to remove this residual field, allowing measurement of periodic moments with no appreciable offset (Figure \ref{newAC}a--c). To optimize sequence run times, magnetic field cycle counts were scaled with frequency (e.g. 20 cycles were measured at 0.0264 Hz and 1 cycle was measured at 0.00013 Hz). An example truncated MPMS3 sequence file can be found in Listing 1 in the SI. The discrete Fourier transform was applied to $H$ vs. $t$ and $M$ vs. $t$ data which yielded complex-valued frequency space representations. Plots of the absolute value of the transform vs. frequency revealed spectra with a primary peak at the fundamental drive field frequency followed by overtone peaks of diminishing intensity. We set $\chi = |FT_{D}(M_{max})|/|FT_{D}(F_{max})$, where $|FT_{D}(M_{max})|$ and $|FT_{D}(F_{max})|$ are absolute values at the fundamental frequency of the complex moment and field spectra, respectively. The in-phase ($\chi{}\prime$) and out-of-phase ($\chi{}\prime\prime$) components of the magnetic susceptibility were then calculated as $\chi{}\prime = \chi{}cos(\phi{})$ and $\chi{}\prime\prime = \chi{}sin(\phi{})$, with $\phi{}$ being the phase angle between the field and moment spectra at the fundamental frequency. Using this method, complex susceptibilities were extracted from VSM moment waveforms of \textbf{1} collected at temperatures between 2--10 K using square-wave drive fields with periods between 38--7690 s.

\begin{figure}
\centering
\includegraphics[width=0.5\linewidth]{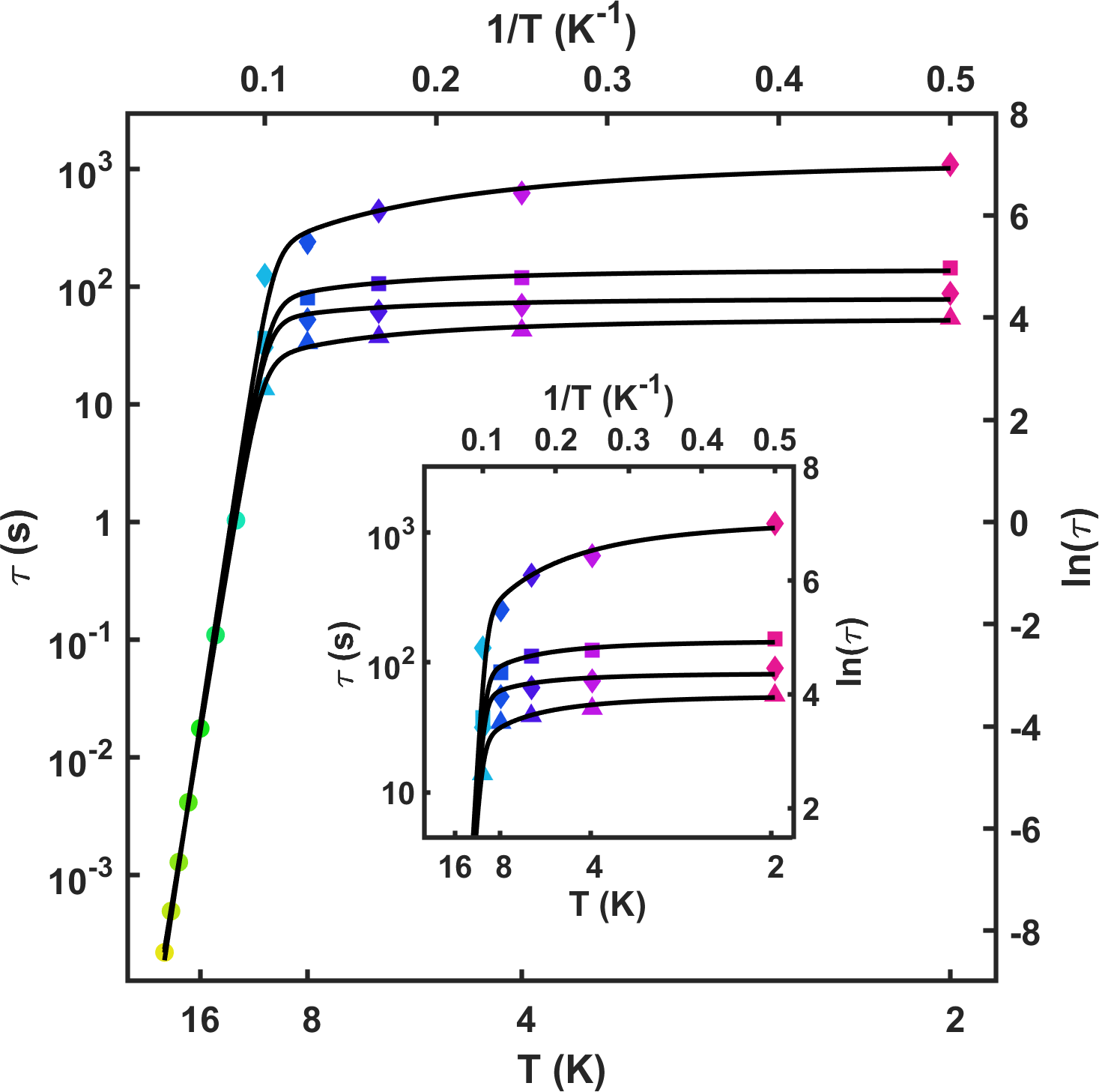}
\caption{Standard AC relaxation times (circles), VSM waveform Debye fit (triangles), stretched exponential fit (squares), and bi-exponential fit (diamonds) of magnetic relaxation data for \textbf{1} with fit to relaxation model S1 (black line, parameters in Table S3). (Inset) Plot of $\tau{}$ vs. 1/\textit{T} showing region of crossover between fitting methods. Black line again shows the relaxation model fit.}
\label{arrhenius}
\end{figure}

Susceptibility parameters extracted from Fourier analysis of the VSM waveform data (Figure \ref{ac_combined}) show a clear phase dependence on drive field frequency. In an approach analogous to that taken with standard AC susceptibilities, we fit in-phase and out-of-phase components of this data to a generalized Debye model (Equations 1 \& 2, Table \ref{fit_values}) and obtained $\alpha{}$ values close to 0 indicating that only a single characteristic relaxation time exists at each temperature. An Arrhenius plot of fitted relaxation times (Figure \ref{arrhenius}) shows that $\tau{}$ is weakly temperature dependent between 2--8 K and at 10 K $\tau{}$ begins to drop considerably as it approaches the high-temperature Orbach relaxation regime. We note that the relaxation time measured at 10 K via this method nearly matches the relaxation time predicted by an Arrhenius fit to the AC susceptibility data. This provides a strong indication that these methods are complementary probes of molecular magnetic relaxation for \textbf{1}. A comparison between relaxation times extracted from VSM waveforms and DC relaxation measurements reveal that each fitting method yields relaxation time profiles that deviate sublinearly from Arrhenius behavior. However, $\tau{}$ values obtained from exponential fits of DC data across the measured temperature range are consistently higher than the corresponding VSM waveform measurements. If we assume the Debye analysis to provide the most accurate $\tau{}$, DC relaxation times extracted with Equation 3 ($\tau_{str}$) and Equation 4 ($\tau_{2}$) represent average errors of 270\% and 175\%, respectively. The long component of Equation 4, $\tau{}_{1}$ represents an even more extreme error of 1261\%. 

\section*{Conclusions}
In this work we have utilized the slow magnetic relaxation behavior of \textbf{1} to test the efficacy of a new technique for extracting characteristic relaxation times that are beyond the scope of standard AC susceptibility experiments. This proof-of-concept work has shown that the VSM waveform method is viable and yields results that are a consistent continuation of the fast timescale dynamics. We hope that given the wide availability of the required instrumentation and the current interest in understanding and controlling the rate of magnetic relaxation mechanisms access to another characterization tool will prove useful. In addition to the expanded timescale, the methods described herein may prove useful for targeting specific relaxation mechanisms that occur at non-zero field. For example, with the field tuned to a level-crossing, quantum-tunneling transitions could be studied in isolation from competing relaxation mechanisms. 
\section*{Acknowledgements}
This research was funded through the Office of Naval Research Young Investigator Award N00014-16-1-2917. The authors thank Drs. Milan Gembicky and Anthony L. Spek (crystallography) and Dr. Anthony Mrse (NMR) for their expert assistance. 

\putbib[bu1]
\end{bibunit}

\clearpage
\renewcommand{\thefigure}{S\arabic{figure}}
\renewcommand{\thetable}{S\arabic{table}}
\renewcommand{\theequation}{S\arabic{equation}}
\setcounter{figure}{0}   
\setcounter{table}{0}   
\setcounter{equation}{0}   
\begin{bibunit}
\begin{center}
    \huge{Supporting Information}
    
\end{center}

\tableofcontents
 
\clearpage
\section{Preparative Details}
\subsection{General Considerations}
Manipulations involving the synthesis of \textbf{1} were carried out in a nitrogen-atmosphere glovebox. Tetrahydrofuran (THF) and pentane were dried on activated alumina columns and stored over a 1:1 mixture of 3 and 4 \angstrom\phantom{a}molecular sieves. 1,6-heptadiyne (purchased from Alfa Aesar) was degassed using freeze-pump-thaw cycles prior to use. Nickel(II) bromide DME complex (Combi-Blocks), 18-crown-6 (18-c-6, Aldrich), anhydrous erbium trichloride (Aldrich), and potassium graphite (KC\textsubscript{8}, Strem) were used as received. THF-d\textsubscript{8} was purchased from Sigma-Aldrich and was degassed using freeze-pump-thaw cycles, dried over stage 0 NaK on silica, and filtered before use. Neutral proligand hdcCOT was synthesized according to a literature procedure.\cite{wender_nickel0-catalyzed_2007-1} \textsuperscript{1}H NMR spectra were collected at --20 \degree{}C on a Jeol ECA 500 spectrometer. CHN elemental analysis was conducted by Midwest Microlab, Indianapolis, IN.

\subsection{\texorpdfstring{[K(18-c-6)][Er(hdcCOT)\textsubscript{2}]}{[K(18-c-6)][Er(hdcCOT)2]} (\textbf{1})}
To a $-50$ \degree{}C stirring suspension of KC\textsubscript{8} (0.368 g, 2.72 mmol, 4 mL THF) was added a solution of hdcCOT (0.239 g, 1.30 mmol, 6 mL THF). The reaction mixture was allowed to warm to room temperature and was stirred for a total of 16 hours. Graphite was removed \textit{via} centrifugation and subsequent filtration through a glass-fiber filter yielded a dark brown homogeneous solution. Concentration \textit{in vacuo} and cooling to $-50$ \degree{}C yielded large yellow plates of K\textsubscript{2}hdcCOT over the course of 24 hours. The mother liquor was removed from these plates and they were subsequently dried \textit{in vacuo}; mass obtained: 0.261 g (Yield: 76.5\%). K\textsubscript{2}hdcCOT (0.182 g, 0.693 mmol) and ErCl\textsubscript{3} (0.095 g, 0.347 mmol) were combined in 20 mL THF and stirred for 16 hours. The resulting yellow suspension was centrifuged and the supernatant was separated and dried \textit{in vacuo} to yield 0.149 g (74.7\%) of K[Er(hdcCOT)\textsubscript{2}] as a bright yellow powder. 18-crown-6 (0.0635 g, 0.240 mmol) was combined with this powder (0.1382 g, 0.240 mmol) in 20 mL THF and the resulting solution was stirred for 16 hours. Vapor diffusion of pentane into this solution yielded yellow rods of \textbf{1} (200 mg, Yield: 85.0\%). CHN analysis (calculated, found) for C\textsubscript{40}H\textsubscript{56}ErK: C (57.25, 54.26); H (6.73, 6.51); N (0, 0).

\clearpage
\section{Sample Characterization}
\subsection{\texorpdfstring{\textsuperscript{1}H NMR of K[Er(hdcCOT)\textsubscript{2}]}{1H NMR of K[Er(hdcCOT)2]}}

\begin{figure}[h]
    \centering
    \includegraphics[width=0.7\linewidth]{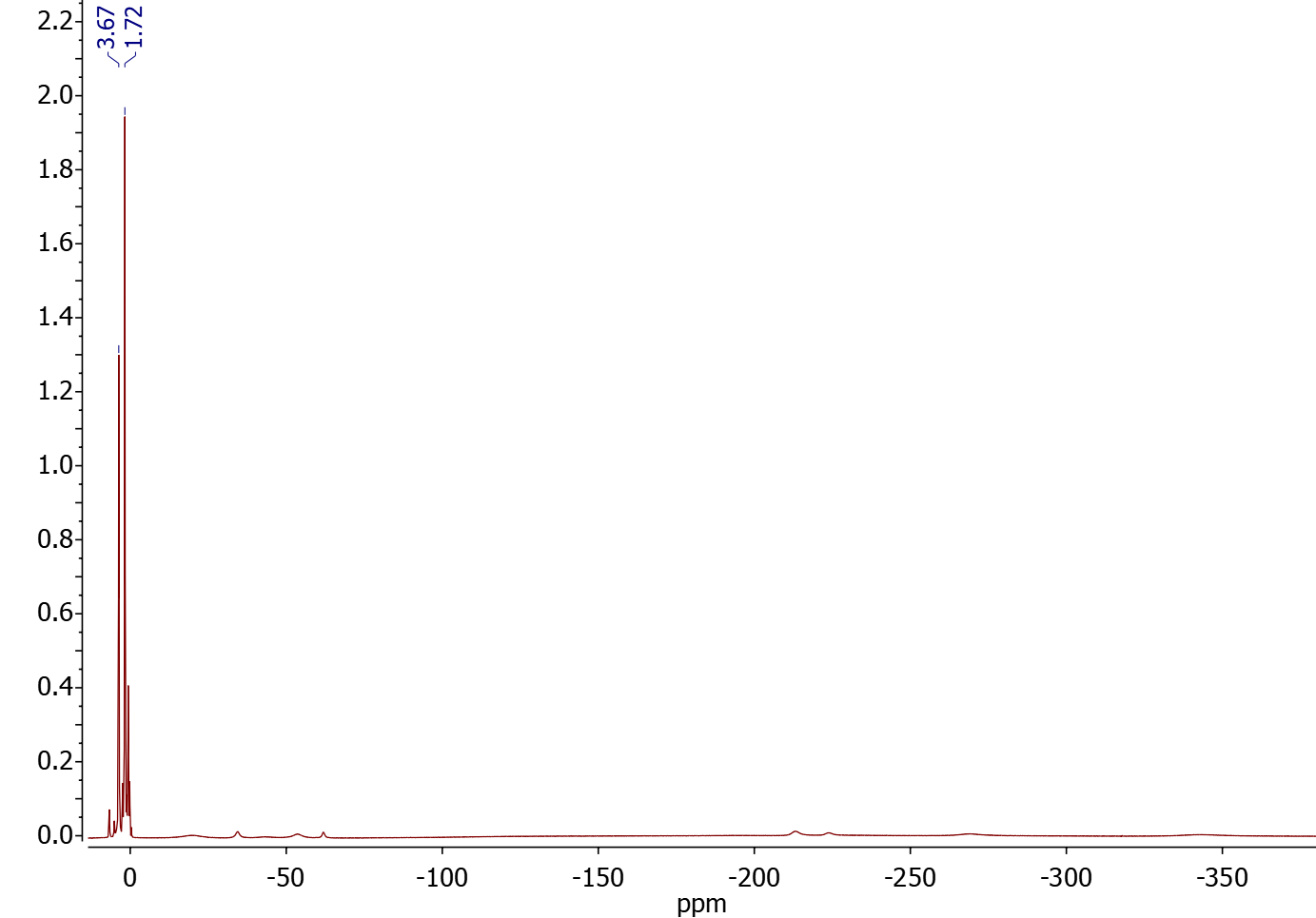}
    \caption{\textsuperscript{1}H NMR spectrum of K[Er(hdcCOT)\textsubscript{2}] in THF-d\textsubscript{8} at --20 \degree{}C. Labelled signals correspond to THF residual peaks.}
    \label{fig:nmr_full}
\end{figure}

\begin{figure}[h]
    \centering
    \includegraphics[width=0.7\linewidth]{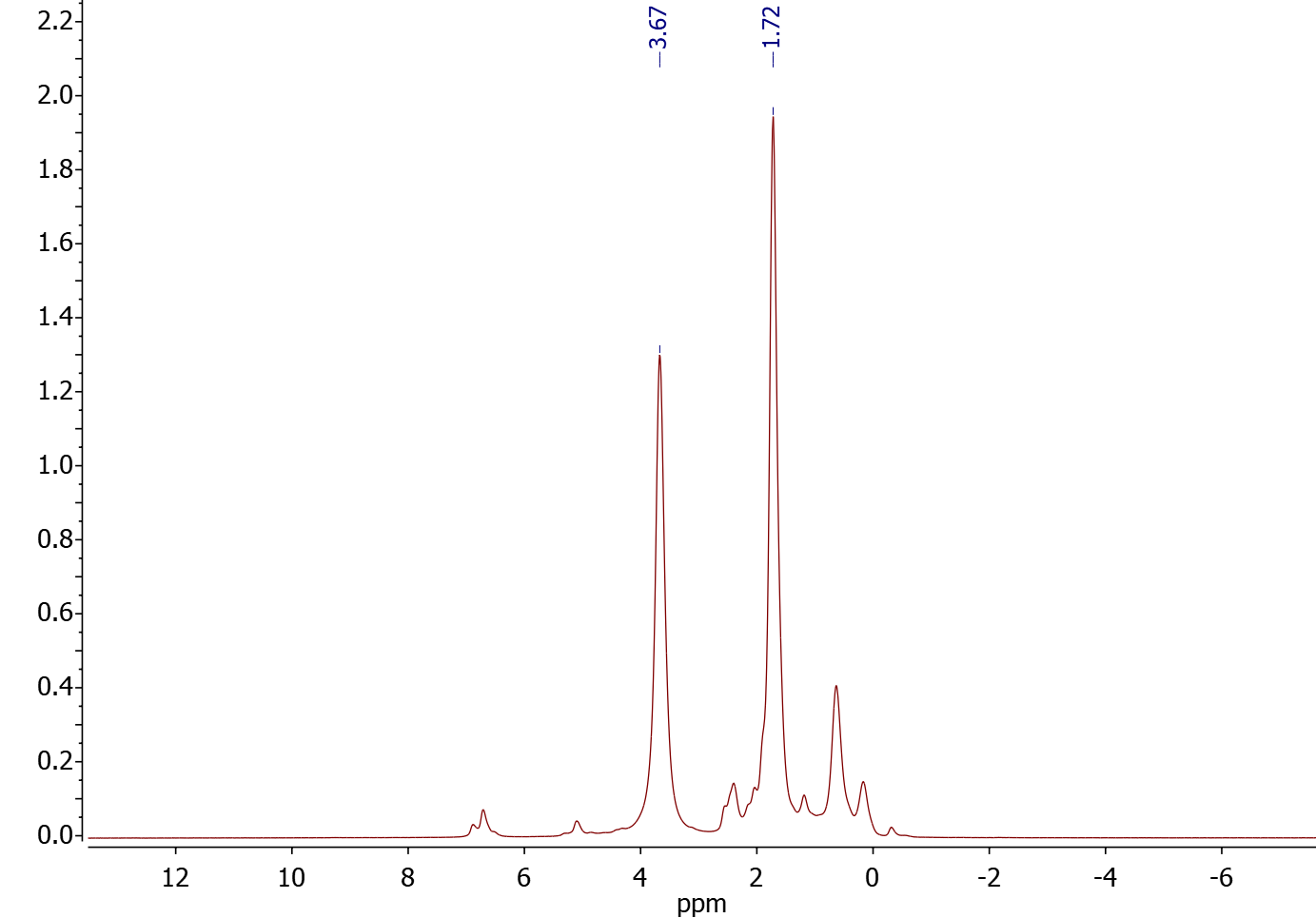}
    \caption{Downfield \textsuperscript{1}H NMR spectrum of K[Er(hdcCOT)\textsubscript{2}] in THF-d\textsubscript{8} at --20 \degree{}C. Labelled signals correspond to THF residual peaks. Unlabelled signals are taken to be decomposition products.}
    \label{fig:nmr_down}
\end{figure}

\begin{figure}[h]
    \centering
    \includegraphics[width=0.7\linewidth]{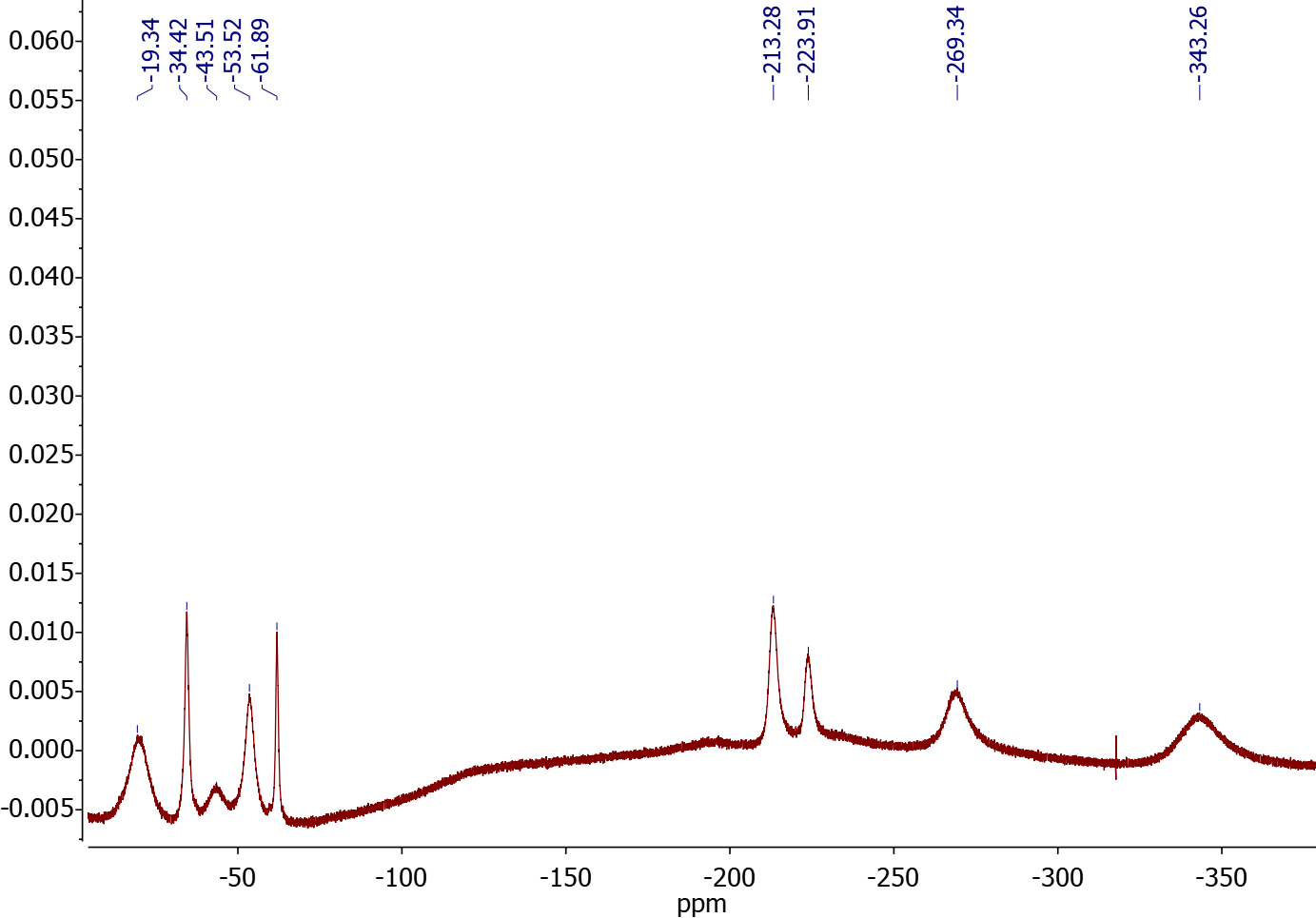}
    \caption{Upfield \textsuperscript{1}H NMR spectrum of K[Er(hdcCOT)\textsubscript{2}] in THF-d\textsubscript{8} at --20 \degree{}C. Labelled signals correspond hdcCOT protons.}
    \label{fig:nmr_up}
\end{figure}
\clearpage

\subsection{Crystallographic Methods}
Single crystal diffraction data for \textbf{1} was collected at 100 K on a Bruker $\kappa$ Diffractometer using a Ga(K$_{\alpha}$) METALJET source and a PHOTON II Area Detector. Data integration was carried out using SAINT and output intensities were corrected for Lorentz and air absorption effects. Additional absorption corrections were applied using SADABS. The structure was solved in space group No. 14 (P2\textsubscript{1}/n) using direct methods with the SHELXT\cite{sheldrick_shelxt_2015} program and anisotropic atom positions were refined against \textit{F\textsuperscript{2}} data using the SHELXL\cite{sheldrick_crystal_2015} program. Olex\textsuperscript{2} was used during the refinement stage as a graphical front-end.\cite{dolomanov_olex2_2009} Real and imaginary anomalous dispersion coefficients for Ga(K$_{\alpha}$) radiation were taken from the Brennan and Cowan\cite{brennanSuiteProgramsCalculating1992} model. The position of all hydrogen atoms were determined using a riding model. Supplementary crystallographic data can be accessed from the Cambridge Crystallographic Data Center, CCDC 1939762.

\subsection{Computational Details}
\textit{Ab initio} electronic structure modelling was carried out at the CASSCF level using the MOLCAS 8.2 software suite. Input atom coordinates were taken from crystallographic data and used without further geometry optimization. K, 18-c-6, and solvent THF were excluded from the input geometry. Basis functions of the ANO-RCC type were generated with the SEWARD module and the quality of a specific atomic basis function was determined as a function of the atom's distance from the Er\textsuperscript{3+} ion (\textbf{Er:} ANO-RCC-VTZP; \textbf{atoms bound to Er:} ANO-RCC-VDZP; \textbf{all other atoms:} ANO-RCC-VDZ). To save disk space and reduce calculation cost two-electron integrals were Cholesky decomposed ($10^{-6}$ cutoff). A 7-orbital, 11-electron activate space (CAS(11,7)) was selected for the CASSCF calculation which was carried out with the RASSCF module. In this space all 35 configuration-interaction (CI) roots of spin multiplicity 4 and all 112 CI roots of spin multiplicity 2 were included. Spin-orbit matrix elements between CAS output wavefunctions were calculated with the RASSI module. SINGLE\_ANISO was used to calculate relevant magnetic properties based on these multiconfigurational SCF results.

\begin{figure}[h]
    \centering
    \includegraphics[width=0.5\linewidth]{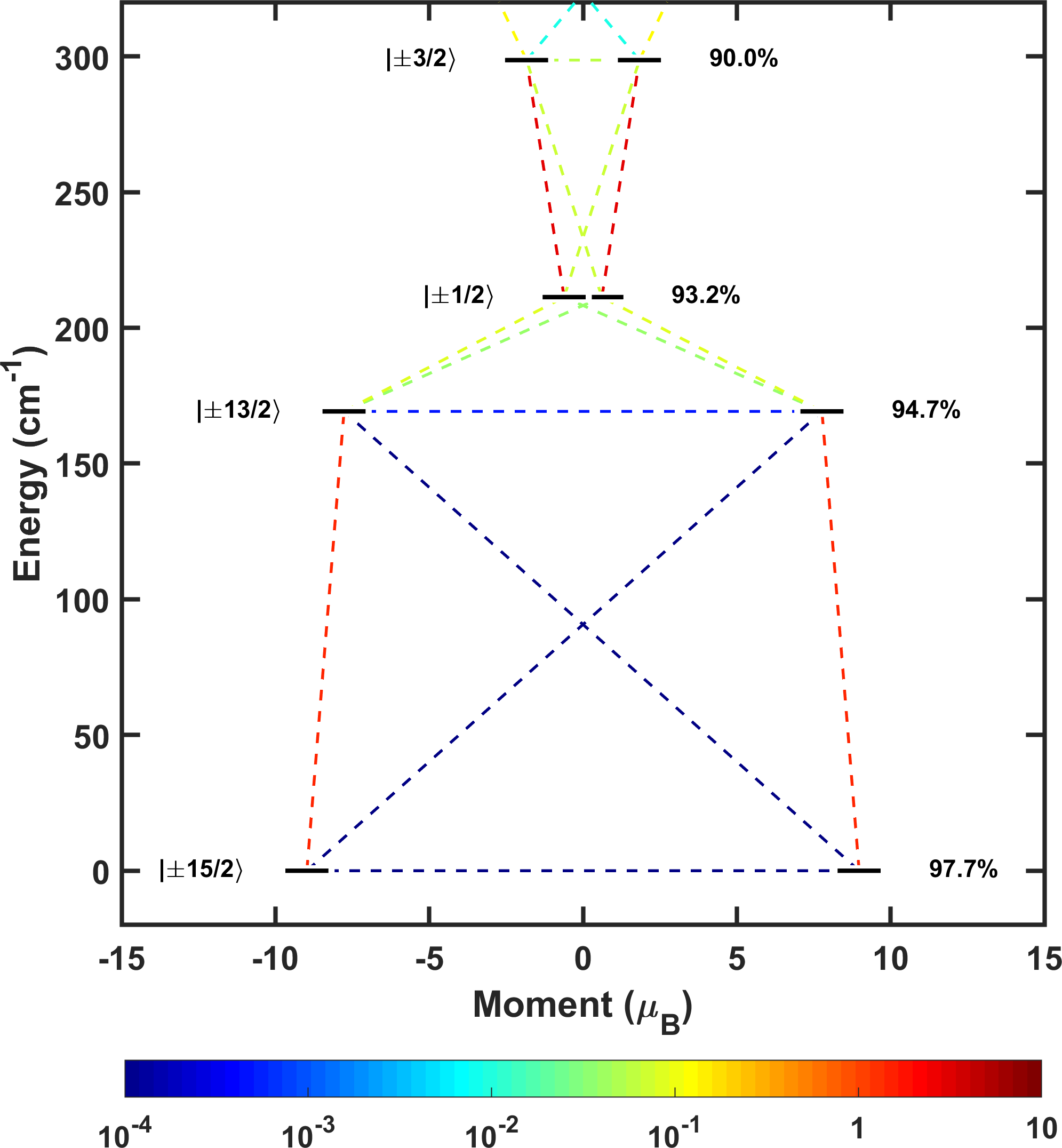}
    \caption{Calculated spectrum of the four lowest energy Kramers states for \textbf{1}. States (black lines) are labelled by their largest $\pm{}m_{J}$ component (left) and the percentage of that component (right). Transverse magnetic moment matrix elements (dotted lines) are colored according to their respective values (colorbar).}
    \label{fig:barrier}
\end{figure}

\begin{table}[h]
\caption{$J = 15/2$ manifold energy spectrum of \textbf{1}.}
\label{state_energies}
\centering
\begin{tabular}{c c c}
\toprule
{KD (i)} & {M\textsubscript{z} ($\mu{}_{B}$)} &  {$E$ (cm\textsuperscript{--1})} \\ 
\midrule
G & 8.976951952 & 0.000 \\ 
1 & 7.769140677 & 169.229 \\ 
2 & 0.613462182 & 211.322 \\ 
3 & 1.834519491 & 298.637 \\ 
4 & 6.547320548 & 411.560 \\ 
5 & 3.031744486 & 433.319 \\ 
6 & 4.722208148 & 542.787 \\ 
7 & 5.730749227 & 550.231 \\
\bottomrule
\end{tabular} 
\end{table}

\begin{table}[h]
\caption{Selected average magnetic moment matrix elements between the $J = 15/2$ multiplets of \textbf{1}.}
\label{matrix_elements}
\centering
\begin{tabular}{c c c c}
\toprule
{KD (i)} & {$\langle{}i_{\uparrow}|\hat{M}_{avg}|i_{\downarrow}\rangle{}$} &  {$\langle{}i_{\uparrow}|\hat{M}_{avg}|i_{\uparrow} + 1\rangle{}$} &  {$\langle{}i_{\uparrow}|\hat{M}_{avg}|i_{\downarrow} + 1\rangle{}$} \\ 
\midrule
G & 0.327451096777E-05 & 0.155011707285E+01 & 0.131694298009E-04 \\ 
1 & 0.506016013299E-03 & 0.928060135350E-01 & 0.405995456082E-01 \\ 
2 & 0.318682266522E+01 & 0.316661251988E+01 & 0.720938920092E-01 \\ 
3 & 0.600396549586E-01 & 0.134946515237E+00 & 0.941722228670E-02 \\ 
4 & 0.142404299502E-02 & 0.242852634761E+00 & 0.973779023853E-02 \\ 
5 & 0.485270066726E-01 & 0.285000005858E+01 & 0.300470628704E-01 \\ 
6 & 0.549287282880E-01 & 0.266558195629E+01 & 0.429984200984E-01 \\ 
7 & 0.508798096471E-01 & & \\
\bottomrule
\end{tabular} 
\end{table}
\clearpage
\subsection{Magnetic Data Collection}
Magnetic data were collected under DC scan and VSM scan modes using a Quantum Design MPMS 3 SQUID Magnetometer with equipped AC susceptibility attachment. Samples were loaded in custom quartz tubes (D\&G Glassblowing Inc.) which were subsequently flame-sealed under static vacuum. To all samples was added a portion of melted eicosane wax to abate sample torquing and facilitate thermal conductivity. Diamagnetic contributions from the sample and eicosane were subtracted from all static moment data using Pascal's constants.\cite{bain_diamagnetic_2008} Magnetic relaxation data were collected in DC scan mode after first equilibrating the sample at a given temperature to a 7 T field then ramping the field (700 Oe sec\textsuperscript{--1}) to 0 T. Details related to the Fourier analysis of long-timescale magnetic data are discussed in the main text. MPMS3 data parsing, fitting, and plotting was performed with MATLAB; the object-oriented code package and documentation used for all processes is available at \url{https://www.github.com/RinehartGroup/qdsquid-dataplot}.

\begin{equation}
\tau^{-1} = \tau_{0}^{-1}exp\left(\frac{-U_{eff}}{k_{B}T}\right) + \tau_{qtm}^{-1} + CT^{2}
\end{equation}
\centering
Relaxation mechanism equation

\clearpage

\begin{lstlisting}[caption=Truncated MPMS3 MultiVu Sequence, basicstyle=\ttfamily\small, numbers=left]
Set Temperature 50K at 50K/min, Fast Settle
Wait For Temperature, Delay 0 sec, No Action

Magnet Reset
Wait For Delay 600 secs (10.0 mins), No Action

Set Temperature 2K at 50K/min. Fast Settle
Wait For Temperature, Delay 120 secs (2.0 mins), No Action

! REMARK - Frequency ~ 0.0264 Hz
MPMS3 Measure for 0.5 sec at 1 mm  every 0 sec Auto-Tracking
Wait For Delay 2 secs, No Action
Scan Time 0.0 secs in 20 steps
    Set Magnetic Field 8.0Oe at 30.00Oe/sec, Linear, Stable
    Wait For Field, Delay 15 secs, No Action
    Set Magnetic Field -8.0Oe at 30.00Oe/sec, Linear, Stable
    Wait For Field, Delay 15 secs, No Action
End Scan
Set Magnetic Field 0.0Oe at 30.00Oe/sec, Linear, Stable
Wait For Field, Delay 0 secs, No Action
Stop Measurements

Wait For Delay 60 secs (1.0 mins), No Action

! REMARK - Frequency ~ 0.0147 Hz
MPMS3 Measure for 0.5 sec at 1 mm  every 0 sec Auto-Tracking
Wait For Delay 2 secs, No Action
Scan Time 0.0 secs in 16 steps
    Set Magnetic Field 8.0Oe at 30.00Oe/sec, Linear, Stable
    Wait For Field, Delay 22 secs, No Action
    Set Magnetic Field -8.0Oe at 30.00Oe/sec, Linear, Stable
    Wait For Field, Delay 22 secs, No Action
End Scan
Set Magnetic Field 0.0Oe at 30.00Oe/sec, Linear, Stable
Wait For Field, Delay 0 secs, No Action
Stop Measurements

! REMARK - Frequencies truncated

Wait For Delay 240 secs (4.0 mins), No Action

! REMARK - Frequency ~ 0.00013 Hz
MPMS3 Measure for 0.5 sec at 1 mm  every 0 sec Auto-Tracking
Wait For Delay 2 secs, No Action
Set Magnetic Field 8.0Oe at 30.00Oe/sec, Linear, Stable
Wait For Field, Delay 3840 secs (1.1 hours), No Action
Set Magnetic Field -8.0Oe at 30.00Oe/sec, Linear, Stable
Wait For Field, Delay 3840 secs (1.1 hours), No Action
Set Magnetic Field 0.0Oe at 30.00Oe/sec, Linear, Stable
Wait For Field, Delay 0 secs, No Action
Stop Measurements

Set Temperature 300K at 50K/min. Fast Settle
Wait For Temperature, Delay 1 secs, No Action
\end{lstlisting}
\clearpage

\begin{table}[h]
\caption{Fitting values for relaxation mechanism parameters. Note these parameters are meant to highlight differences in the form of the relaxation data determined by different methods and are purely phenomenological.}
\label{relaxation_parameters}
\centering
\begin{tabular}{c c c c c}
\toprule
{} & {$U_{eff}$ (cm\textsuperscript{--1})} &  {$\tau_{0}$ (s)} &  {$\tau_{qtm}$ (s)} & {C (K\textsuperscript{--2}s\textsuperscript{--1})} \\ 
\midrule
$\tau_{AC}$ & 147.7(7) & 3.1(1)$\times$10\textsuperscript{--8} &  & \\ 
$\tau_{AC} + \tau_{VSM}$ & 142(3) & 4.7(1.0)$\times$10\textsuperscript{--8} & 54(7) & 2.2(8)$\times$10\textsuperscript{-4}\\ 
$\tau_{AC} + \tau_{str}$ & 146(5) & 3.5(1.1)$\times$10\textsuperscript{--8} & 141(31) & 6.2(6.2)$\times$10\textsuperscript{--5}\\ 
$\tau_{AC} + \tau_{1}$ & 154(13) & 1.8(1.0)$\times$10\textsuperscript{--8} & 1209(621) & 4.0(4.2)$\times$10\textsuperscript{--5}\\ 
$\tau_{AC} + \tau_{2}$ & 145(6) & 3.6(1.2)$\times$10\textsuperscript{--8} & 80(17) & 6.9(9.7)$\times$10\textsuperscript{--5}\\
\bottomrule
\end{tabular} 
\end{table}

\begin{figure}[h!]
    \centering
    \includegraphics[width=0.55\linewidth]{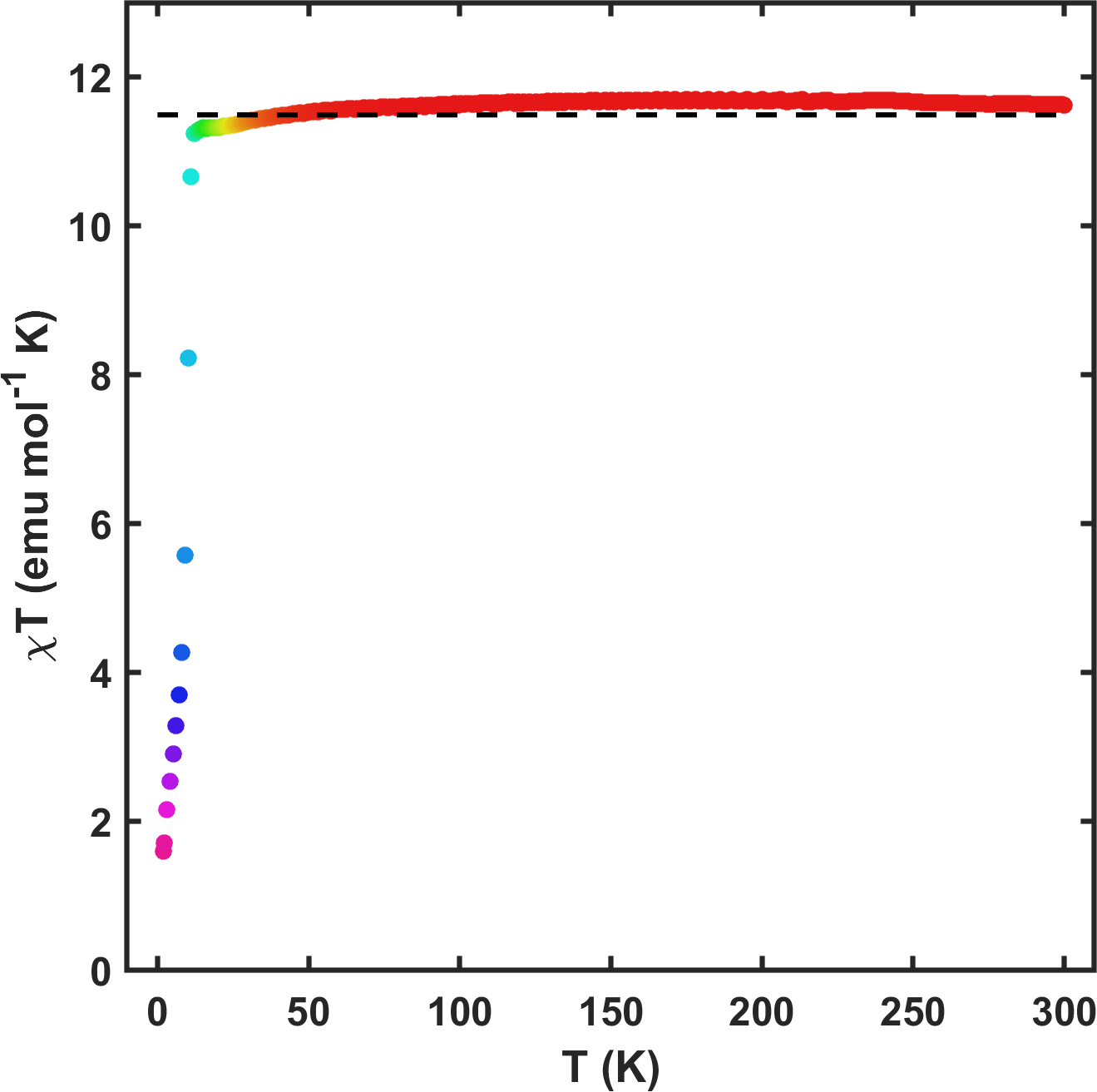}
    \caption{Magnetic susceptibility of \textbf{1} between 2 to 300 K. Colored dots are data measured under a 1000 Oe field and the dotted line is the theoretical $\chi{}T$ value for a free Er\textsuperscript{3+} ion.}
    \label{fig:susceptibility}
\end{figure}

\begin{figure}[h!]
    \centering
    \includegraphics[width=0.55\linewidth]{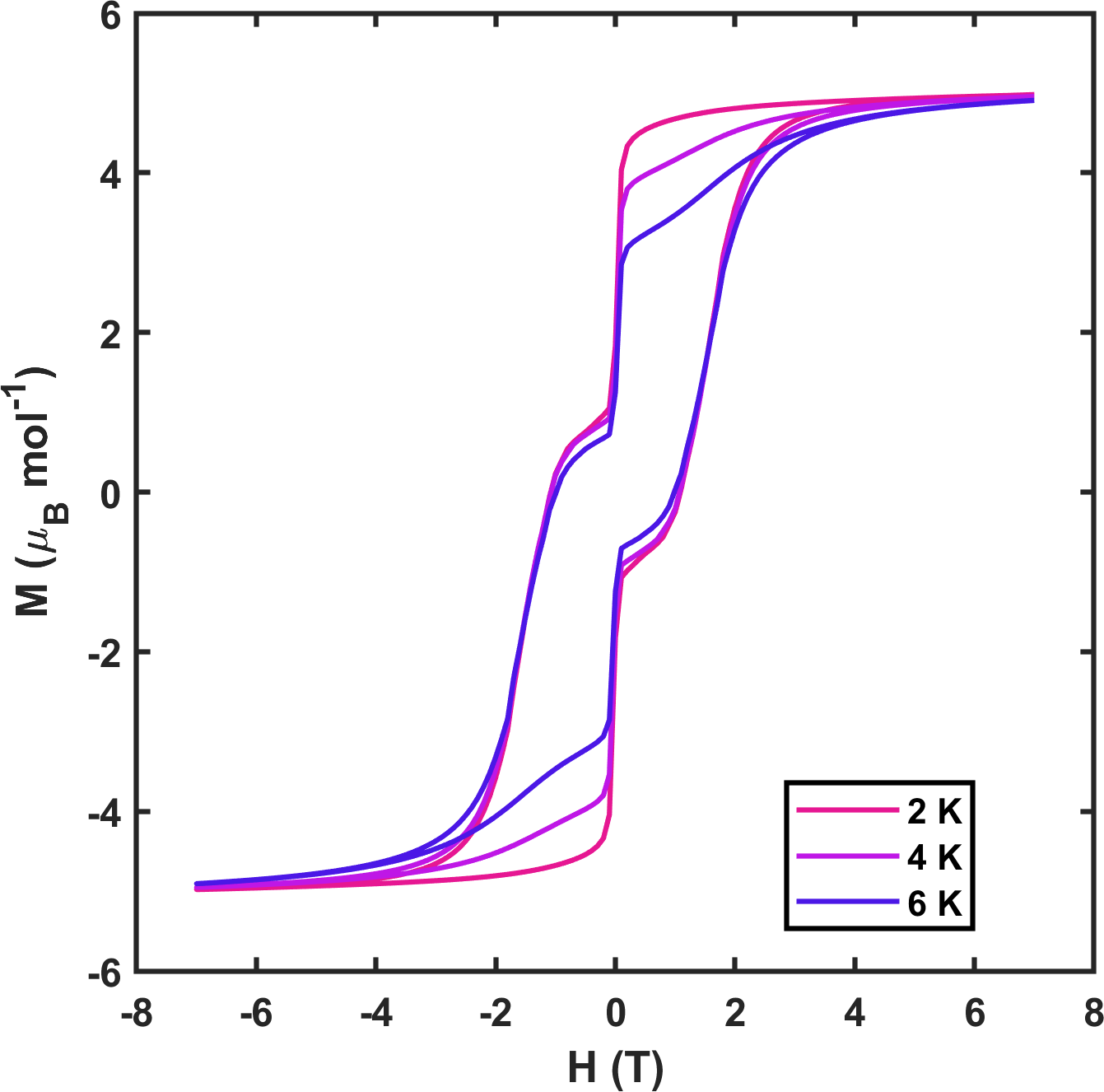}
    \caption{Isothermal magnetization of \textbf{1} from --7 to 7 T (Rate = 10 Oe sec\textsuperscript{--1}).}
    \label{fig:magnetization}
\end{figure}

\begin{figure}[h!]
    \centering
    \includegraphics[width=0.55\linewidth]{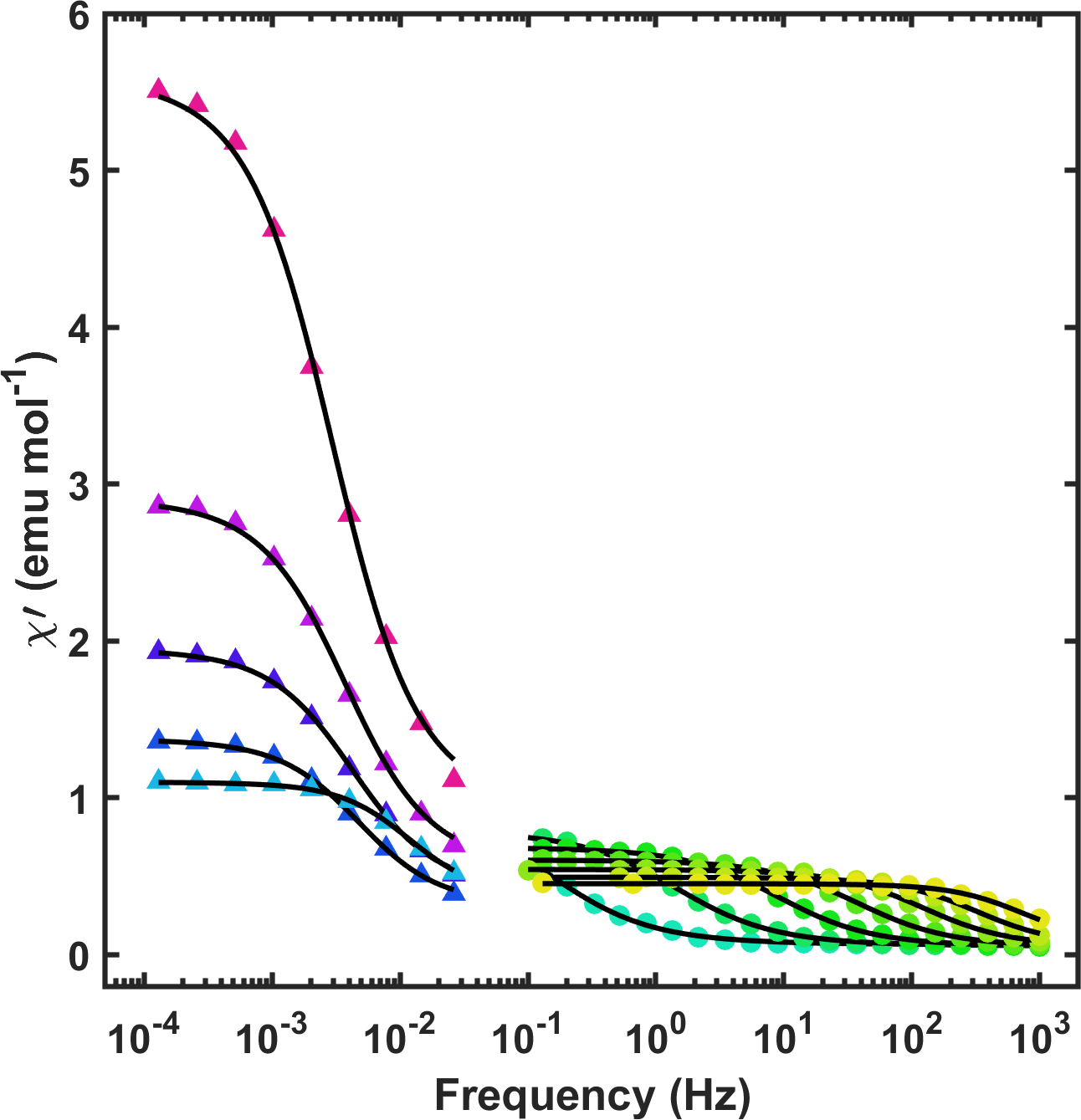}
    \caption{In-phase magnetic susceptibility of \textbf{1} bewteen 2 (fuschia) and 24 (yellow) K. Colored points are susceptibilities measured via standard AC measurements (circles) and extracted from Fourier analysis of VSM data (triangles). Black lines represent fits to a generalized Debye model (Equations 1 \& 2).}
    \label{fig:in_phase}
\end{figure}

\begin{figure}[h!]
    \centering
    \includegraphics[width=0.55\linewidth]{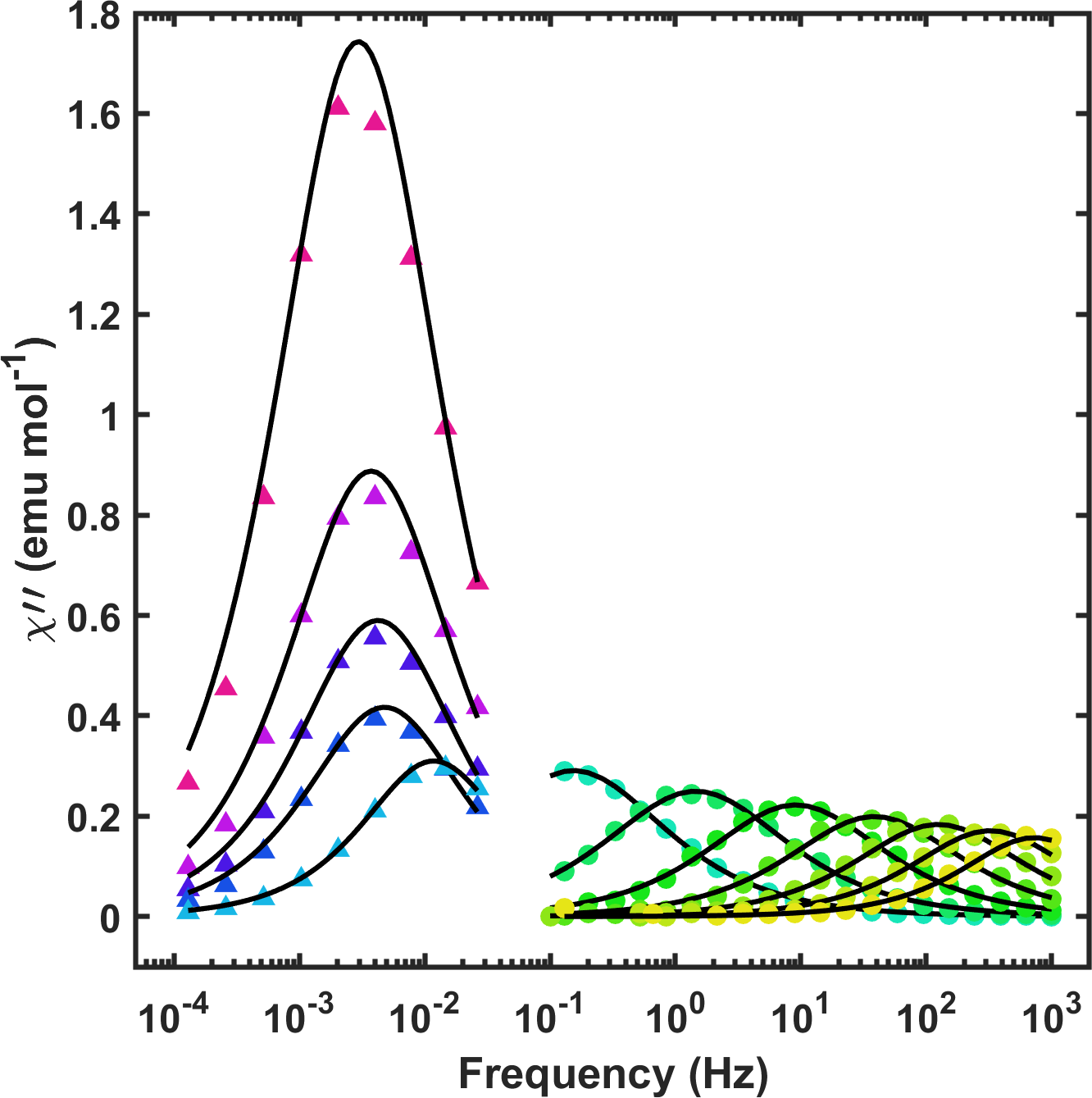}
    \caption{Out-of-phase magnetic susceptibility of \textbf{1} bewteen 2 (fuschia) and 24 (yellow) K. Colored points are susceptibilities measured via standard AC measurements (circles) and extracted from Fourier analysis of VSM data (triangles). Black lines represent fits a to generalized Debye model (Equations 1 \& 2).}
    \label{fig:out_of_phase}
\end{figure}
\clearpage
\addcontentsline{toc}{section}{References}
\putbib[bu2]
\end{bibunit}

\end{document}